\documentclass{aa}  

\usepackage{graphicx}
\usepackage{txfonts}
\usepackage{soul}
\usepackage{color}

\usepackage{mathabx}

\usepackage{natbib,twoopt}
\usepackage[breaklinks=true]{hyperref} 
\bibpunct{(}{)}{;}{a}{}{,}             
\makeatletter
  \newcommandtwoopt{\citeads}[3][][]{\href{http://adsabs.harvard.edu/abs/#3}%
    {\def\hyper@linkstart##1##2{}%
     \let\hyper@linkend\@empty\citealp[#1][#2]{#3}}}
  \newcommandtwoopt{\citepads}[3][][]{\href{http://adsabs.harvard.edu/abs/#3}%
    {\def\hyper@linkstart##1##2{}%
     \let\hyper@linkend\@empty\citep[#1][#2]{#3}}}
  \newcommandtwoopt{\citetads}[3][][]{\href{http://adsabs.harvard.edu/abs/#3}%
    {\def\hyper@linkstart##1##2{}%
     \let\hyper@linkend\@empty\citet[#1][#2]{#3}}}
  \newcommandtwoopt{\citeyearads}[3][][]%
    {\href{http://adsabs.harvard.edu/abs/#3}
    {\def\hyper@linkstart##1##2{}%
     \let\hyper@linkend\@empty\citeyear[#1][#2]{#3}}}
\makeatother

\usepackage{dcolumn}

\newcolumntype{K}{D{.}{.}{2,2}}
\newcolumntype{H}{D{,}{\pm}{4.4}}

\newcommand\N{21}

\newcommand\Nb{17}

\begin{document}

\title{VLA proper motion constraints on the origin, age, and potential magnetar future of PSR J1734$-$3333}


%
%
%
%
%

\titlerunning{The proper motion of PSR J1734$-$3333}

\author{C.~M. Espinoza\inst{1,2}
	\and
	M. Vidal-Navarro\inst{3}
	\and 
	W.~C.~G. Ho\inst{4}
	\and
	A. Deller\inst{5}
	\and
	S. Chatterjee\inst{6}
}

\institute{Departamento de F\'isica, Universidad de Santiago de Chile (USACH), Av. Victor Jara 3493, Estaci\'on Central, Chile \\
	\email{cristobal.espinoza.r@usach.cl}
	\and
	Center for Interdisciplinary Research in Astrophysics and Space Sciences (CIRAS), Universidad de Santiago de Chile, Chile 
	\and
	N\'ucleo de Astroqu\'imica, Facultad de Ingenier\'ia, Universidad Aut\'onoma de Chile, Av. Pedro de Valdivia 425, Providencia, Santiago, Chile
	\and
	Department of Physics and Astronomy, Haverford College, 370 Lancaster Avenue, Haverford, PA, 19041, USA
	\and
Centre for Astrophysics and Supercomputing, Swinburne University of Technology, John St, Hawthorn, VIC 3122, Australia
	\and
	Department of Astronomy and Cornell Center for Astrophysics and Planetary Science, Cornell University, Ithaca, New York, 14853, USA
}

\date{}

\abstract
{The characteristic age of PSR J1734$-$3333 estimated from its current spin down rate implies that it is a young pulsar ($\tau_c<10$\,kyr). 
But the time derivative of its spin down rate differs markedly from that assumed for normal radio pulsars, meaning its actual age is uncertain.
G354.8$-$0.8 is a supernova remnant (SNR) whose centre is located 21\arcmin\, away of the pulsar, and with a morphology that suggests an association with the pulsar.}
{We want to assess the likelihood of the association between PSR J1734$-$3333 and G354.8$-$0.8 or other nearby supernova remnants quantitatively, with the objective of shedding light on the real age of this pulsar. }
{Observations with the Karl G. Jansky Very Large Array were carried out in 2015 and 2019 that allow precise astrometric measurements and consequently a proper motion estimate for the pulsar.}
{The proper motion was found to be $\mu_\alpha=10\pm10$ \,mas\,yr$^{-1}$ and $\mu_\delta=-29\pm11$\,mas\,yr$^{-1}$ (error bars are $1$-$\sigma$). 
Though marginal, this detection rules out the association with G354.8$-$0.8 because it means the pulsar is not moving away from the centre of the SNR. 
No SNR consistent with the measured proper motion and an age $\sim\tau_c$ could be found.
We also present the first measurement of the spectral index for this pulsar, $\alpha=-1.1\pm0.3$, measured between $1.5$ and $3.0$\,GHz.
}
{The SNR produced by the birth supernova of PSR J1734$-$3333 could have already faded to undetectable brightness, for which estimates suggest timescales of $10$-$100$\,kyr. 
This and other considerations lead us to conclude that the pulsar is possibly older than $45$-$100$\,kyr. 
PSR J1734$-$3333 is a pulsar with rotational properties that place it between standard radio pulsars and magnetars, and we interpret our result in the context of a possible future life as a magnetar for this pulsar. 
}


\maketitle

\section{Introduction}
The age of a neutron star is difficult to measure, as for many other astronomical sources.  
The most robust way to do it is by identifying the birth supernova of the neutron star. 
However, this can be done precisely only for a very small number of objects, as 5-10 supernovae have historically been observed in our galaxy \citepads{2005JHA....36..217S}, and neutron stars are faint sources --practically undetectable at distances beyond the Magellanic Clouds.
The explosions, however, leave imprints in the interstellar medium that can remain visible at radio wavelengths for $10-100$\,kyr \citepads{2017MNRAS.464.2326S}, thereby allowing the association of pulsars with supernova remnants (SNRs).
However, pulsars are rarely found at the centre of SNRs \citepads{1994ApJ...437..781F}, as most are expelled like bullets during the explosions possibly due to asymmetries in the process \citepads[e.g.][]{2005ApJ...632..531S}.
The transverse velocities of pulsars (based on proper motion and distance estimates) are particularly large, with a mean close to $310$\,km\,s$^{-1}$ \citepads{2005MNRAS.360..974H}, which is at least 10 times larger than the average velocities for stars in the solar neighbourhood \citepads[e.g.][]{2018A&A...616A..11G}, 
and with some measured velocities ranging as high as $1000$\,km\,s$^{-1}$ \citepads{2005ApJ...630L..61C,2019ApJ...875..100D}.

Thus, associations between SNRs and pulsars are not always straightforward to make \citepads[e.g. see the chapter on young pulsars in][]{2012puas.book.....L}.  
The farther the pulsar is from the explosion site, the higher the possibility that pulsar and SNR are unrelated. 
In order to confirm an association, it could be necessary to account for up to $100$\,kyr of evolution of the SNR (that we assume as the maximum possible age of a SNR), and movement across the Galaxy of the pulsar \citepads[e.g.][]{2021ApJ...914..103S}.
In some situations, proper motion measurements for the pulsars can shed light on the matter. For an association to be secure, the pulsar must be moving away from where the explosion took place (usually adopted as the centre of the SNR), and the time necessary to move the pulsar to its current position must match the age of the system. 
If such time coincided with an independent age measurement of the SNR or the pulsar, or both, then the association would be firmly confirmed. 
But this is rarely possible, as SNR and pulsar ages are hard to obtain.

In this paper we present interferometric observations of PSR J1734$-$3333 designed to measure its proper motion, evaluate its possible connection with the shell-type SNR G354.8$-$0.8 \citepads{1996A&AS..118..329W,2014BASI...42...47G}, and, if possible, constrain the age of the pulsar.  
This association was first proposed by \citetads{2002ASPC..271...31M} based on the proximity of the two objects and the fact that the SNR exhibits a teardrop shape oriented towards the pulsar.
But, given that the evidence is circumstantial and that the pulsar is 21\arcmin\ away of the centre of the SNR, the association was not considered secure \citepads{2002ASPC..271...31M}.
No estimates for the distance of SNR G$354.8$$-$$0.8$ are available.
While confirming this association is in itself important, PSR J1734$-$3333 possesses some other interesting properties, like a decreasing characteristic age and a potential future as a magnetar, which add relevance to the proper motion measurement that we present.

\subsection{The age and braking index of PSR J1734$-$3333}
Another common way to estimate the age of a pulsar is by measuring its period $P$ and period time derivative $\dot{P}$, assume a spin down law and calculate the time required to reach the current value of $P$. 
Typically, the rotation is believed to be slowed down by magnetic dipole radiation, which implies $\dot{P}\,\propto\,P^{-1}$ for a constant dipole magnitude.  
If additionally the initial period $P_0$ is assumed to have been significantly shorter than today's value, that is if $\left(P_0/P\right)^2\ll1$, then the age is given by $\tau_c=P/2\dot{P}$, which is known as the characteristic age \citepads[e.g.][]{1977puls.book.....M,2012puas.book.....L}.
However, the caveats involved can sometimes be enough to make this age definition unreliable.

PSR J1734$-$3333 rotates with a typical period $P=1.169$\,s, but slows down with the fourth largest period derivative amongst pulsars \citepads[$\dot{P}=2.3\times10^{-12}$,][]{2002MNRAS.335..275M,2005AJ....129.1993M}.
These measurements imply the particularly young characteristic age $\tau_c=8.1$\,kyr.
However, a measurement of the second derivative of the spin period
indicates that the rotation of this pulsar may not be driven by
constant magnetic dipole braking.

The braking index ($n$) is used to characterise the long-term
rotational evolution of pulsars, and is defined from the relation
$\dot{P}\,\propto\,P^{2-n}$
\citepads[e.g.][]{1977puls.book.....M,2012puas.book.....L}.
Pulsars evolving purely under the effects of magnetic dipole braking
are expected to exhibit a braking index of $3$.  However, measurements
point to either $n<3$ in very young pulsars like PSR J1734$-$3333 or to
$10\leq n\leq100$, for cases possibly affected by timing noise or
glitch recoveries \citepads{2020MNRAS.494.2012P}.  To measure the
braking index it is necessary to detect the period second derivative,
which is a fairly small quantity, hence difficult to detect
\citepads{2017MNRAS.466..147E}. 
In the case of PSR J1734$-$3333, two independent measurements indicate that $n$ is in the range $0.7$-$1.4$, one of the lowest braking indices ever measured
\citepads[][]{2011ApJ...741L..13E,2021MNRAS.508.3251L}. 
Such low braking index carries some consequences, amongst which is
that the characteristic age is decreasing (if $n<1$), or increasing very slowly
with time \citepads[if $n\gtrsim 1$,][]{2004IAUS..218..257L}.  
Therefore, $\tau_c$ is a highly unreliable proxy for the age of this pulsar, which implies that PSR J1734$-$3333 could be substantially older than $8.1$\,kyr.

\subsection{PSR J1734$-$3333 and the magnetars}
Both $P$ and $\dot{P}$ measurements enable the inference of the amplitude of the effective dipole magnetic field responsible for the slow down, should magnetic dipole radiation be the dominant braking mechanism \citepads[e.g.]{1977puls.book.....M}.  
It corresponds to $B=3.2\times10^{19}(P\dot{P})^{1/2}$, where standard assumptions regarding the orientation of the dipole and the size of the star have been made, and for PSR J1734$-$3333 amounts to $B=5.2\times10^{13}$\,G, which is the fourth largest magnetic field strength known amongst pulsars \citepads[The ATNF Pulsar Catalogue,][]{2005AJ....129.1993M}\footnote{http://www.atnf.csiro.au/research/pulsar/psrcat}, and higher than the inferred values for at least three of the approximately 30 known magnetars \citepads[The McGill Online Magnetar Catalog,][]{2014ApJS..212....6O}\footnote{http://www.physics.mcgill.ca/~pulsar/magnetar/main.html}.
Magnetars are a class of neutron stars, which emit in the X-rays and are identified by their luminous X-ray and soft $\gamma$-ray outburst episodes that usually require more power than is available in the rotation of the star.  
They rotate with periods in the range $2$-$12$\,s, which is slower than most pulsars, and slowdown quicker than all pulsars \citepads[with the notable exception of three sources with   relatively low-$\dot P$ values,][]{2010Sci...330..944R,2012ApJ...754...27R,2014ApJ...781L..17R}).
Both the high $\dot{P}$ values and the high energy emission of magnetars are
believed to be driven and powered by the decay of very intense
magnetic fields of $10^{14-15}$\,G
\citepads[e.g.][]{2006csxs.book..547W,2015SSRv..191..315M,2017ARA&A..55..261K}.
This is in opposition to most other, rotation powered pulsars that
rotate faster, have lower dipole field strengths, exhibit modest and
stable X-ray emission (if any), and are observed mostly in radio
wavelengths.

PSR J1734$-$3333 is a normal radio pulsar with standard X-ray emission
properties, unlike the magnetars \citepads[though there is a debate on whether the pulsar is hotter than others of similar $\tau_c$, see][]{2013ApJ...764....1O,2013ApJ...764..180K}.  
However, its very low braking index implies that its rotational properties are evolving quickly towards those of the magnetars.  
In $\sim30$\,kyr, if the braking index remained low, the pulsar will rotate with a period of $8$\,s and the dipole magnetic field strength $B$ inferred from the pulsar spin-down would be comparable to that for magnetars \citepads{2011ApJ...741L..13E}.

The separation between magnetars and rotation powered pulsars has indeed become less clear as new neutron stars have been discovered \citepads[e.g.][]{2020ApJ...902....1H} and observations of previously known pulsars accumulate.  
High-$B$ rotation powered pulsars have been seen to turn into magnetars for a few months \citepads[see the X-ray outbursts of PSRs J1846$-$0258 and J1119$-$6127,][]{2008Sci...319.1802G,2009A&A...501.1031K,2016ApJ...829L..21A} and some magnetars have been seen to awake in the radio band during or after outbursts \citepads[e.g.][]{2006Natur.442..892C,2007ApJ...666L..93C,2013Natur.501..391E}. 
Therefore, it is possible that some high-$B$ pulsars are dormant magnetars, which any day could display full magnetar high-energy behaviour.
Note that there is not a single rotation powered pulsar known with the rotational properties of magnetars: large periods and period derivatives.  
Thus PSR J1734$-$3333, with its high $\dot P$, could be an evolutionary link between the two classes of pulsars.
Collectively, these facts make its age, proper motion, and relationship with SNR G354.8$-$0.8 particularly relevant.
Evolutionary links between neutron star families are one possible solution to explain the overabundance of neutron stars in the Galaxy with respect to the rate of supernova explosions \citepads{2008MNRAS.391.2009K}.

\section{Data}
We used the VLA on two epochs separated by approximately $4.2$\,yr to
image the field of PSR J1734$-$3333 and constrain its proper motion.

\subsection{Observations}

The first observations were carried out on May 24$^\textrm{th}$, 2015,
using the BnA VLA configuration. This configuration was chosen due to
the low declination ($-33^\circ$) of the pulsar, which at observatory
site translates to a maximum elevation of $\sim 20^\circ$. On this
opportunity the observations were done on a single pointing using both
the L (1--2\,GHz) and S (2--4\,GHz) bands. The second observation was
carried out only at the S band, on August 21$^\textrm{st}$ 2019. After
2015 the BnA configuration was not offered anymore, hence we used the A
configuration for the second observation. Table \ref{tab:log} lists a
summary of the observations.

For both observation epochs, the primary calibrator was 3C286 (flux density and bandpass), and the secondary calibrator was J1713-3226 (for phase). 
The data were calibrated following standard procedures, using the VLA calibration pipeline that is built on the Common Astronomy Software Applications \citepads[CASA][]{2007ASPC..376..127M} versions 4.4.2 and 5.4.2. 
After an initial calibration run, we inspected manually the resulting visibilities and calibration tables in order to search for problematic data (either RFI or instrumental). RFI was present in the S-band data of both observations of both epochs. 
After the removal of the problematic visibilities, we reran the calibration pipeline and iterated the entire process until no signs of bad data were detected. The flagged data in S-band accounts for $\sim\!20\%$ and $\sim\!14\%$ of the on-time source for the 2015 and 2019 data, respectively.

\begin{table*}[!ht]
  \centering
  \caption{Log of observations.}
  \label{tab:log}
  \begin{tabular}{lccccccccc}
    \hline \hline
    Date         & Array$^{(a)}$  & Frequency$^{(b)}$   & Synthesized beam$^{(c)}$      & $\sigma$ \\ 
    &             &   (GHz)           &   (arcsec)        &   ($\mu$Jy\,bm$^{-1}$) \\
    \hline
    24-May-2015    &   BnA       &  1.5            & (5.3 x 2.1 / -48) &  100  \\
    24-May-2015    &   BnA       &  3.0            & (2.5 x 1.4 / -44) &  8.6 \\
    2-May-2019     &   A         &  3.0            & (1.5 x 0.5 / 9.5) &  7.9 \\
    \hline
  \end{tabular}
  \begin{flushleft}
  \tablefoot{
  	\tablefoottext{a}{VLA configuration.}
	\tablefoottext{b}{Centre frequencies for each band that is made up of $2048\times 1$\,MHz channels.}
	\tablefoottext{c}{Briggs-weighted beam in arcsec, in the form (BMAJ x BMIN / BPA), where BMAJ and BMIN are the full-width major and minor axis, and BPA is the beam parallactic angle (PA) in degrees East of North.}
	}
  \end{flushleft}
\end{table*}

\begin{table*}
\centering
\caption{Source parameters from the 2015 map obtained with the CASA routine {\sc imfit}.}
\label{tab:src1}
\begin{tabular}{lHKKccKK}
\hline
\hline
\multicolumn{1}{c}{ID}
& \multicolumn{1}{c}{Flux Density}  
& \multicolumn{1}{c}{S/N}
& \multicolumn{1}{c}{Dist. PSR}
& \multicolumn{1}{c}{RA}  
& \multicolumn{1}{c}{DEC}
& \multicolumn{1}{c}{1-$\sigma$ RA uncertainty}  
& \multicolumn{1}{c}{1-$\sigma$ DEC uncertainty } \\ 
\multicolumn{1}{c}{ }
& \multicolumn{1}{c}{($\mu{\rm Jy}$)}  
& \multicolumn{1}{c}{ }
& \multicolumn{1}{c}{(arcsec)}
& \multicolumn{1}{c}{J2000}  
& \multicolumn{1}{c}{J2000}
& \multicolumn{1}{c}{(mas) }  
& \multicolumn{1}{c}{(mas) }\\
\hline
 3 &  94.4 ,   8.5 &  11.1 &  554.7 &17$^{\rm h}$34$^{\rm m}$31$^{\rm s}\!.$0096 & -33$^{\circ}$42$^{\prime}$32$^{\prime \prime}\!\!.$3247 &   93.0 &  103.8  \\
 4 & 143.3 ,   8.3 &  17.3 &  389.1 &17$^{\rm h}$34$^{\rm m}$19$^{\rm s}\!.$2379 & -33$^{\circ}$39$^{\prime}$37$^{\prime \prime}\!\!.$0915 &   69.9 &   73.0  \\
 5 & 257.4 ,   8.2 &  31.2 &  349.7 &17$^{\rm h}$34$^{\rm m}$33$^{\rm s}\!.$1721 & -33$^{\circ}$39$^{\prime}$00$^{\prime \prime}\!\!.$8382 &   37.3 &   30.6  \\
 7 & 129.5 ,   8.7 &  14.9 &  706.1 &17$^{\rm h}$33$^{\rm m}$36$^{\rm s}\!.$6996 & -33$^{\circ}$38$^{\prime}$44$^{\prime \prime}\!\!.$3096 &   86.6 &   99.4  \\
 8 & 173.6 ,   8.1 &  21.3 &  364.9 &17$^{\rm h}$34$^{\rm m}$04$^{\rm s}\!.$9805 & -33$^{\circ}$37$^{\prime}$21$^{\prime \prime}\!\!.$1721 &   57.7 &   59.0  \\
 9 & 400.5 ,   8.6 &  46.5 &  533.6 &17$^{\rm h}$35$^{\rm m}$05$^{\rm s}\!.$4970 & -33$^{\circ}$37$^{\prime}$08$^{\prime \prime}\!\!.$2245 &   24.8 &   24.0  \\
 10 &  51.4 ,   7.1 &   7.3 &  273.4 &17$^{\rm h}$34$^{\rm m}$41$^{\rm s}\!.$7640 & -33$^{\circ}$36$^{\prime}$40$^{\prime \prime}\!\!.$5500 &  272.7 &  169.7  \\
 11 &  82.5 ,   7.5 &  11.0 &  196.2 &17$^{\rm h}$34$^{\rm m}$29$^{\rm s}\!.$3675 & -33$^{\circ}$36$^{\prime}$33$^{\prime \prime}\!\!.$7612 &   76.2 &   83.5  \\
 12 & 109.6 ,   8.3 &  13.1 &  568.8 &17$^{\rm h}$35$^{\rm m}$09$^{\rm s}\!.$9687 & -33$^{\circ}$36$^{\prime}$23$^{\prime \prime}\!\!.$9391 &   99.2 &   95.4  \\
 13 &  68.0 ,   8.6 &   7.9 &  227.1 &17$^{\rm h}$34$^{\rm m}$38$^{\rm s}\!.$9762 & -33$^{\circ}$36$^{\prime}$09$^{\prime \prime}\!\!.$7422 &  169.6 &   91.6  \\
 14 & 152.3 ,   8.1 &  18.7 &  340.0 &17$^{\rm h}$34$^{\rm m}$00$^{\rm s}\!.$2694 & -33$^{\circ}$34$^{\prime}$29$^{\prime \prime}\!\!.$4742 &   50.7 &   61.7  \\
 17$^{p}$ & 343.7 ,   8.3 &  41.5 &    2.8 &17$^{\rm h}$34$^{\rm m}$27$^{\rm s}\!.$0591 & -33$^{\circ}$33$^{\prime}$21$^{\prime \prime}\!\!.$9157 &   25.8 &   25.9  \\
 18 &  88.6 ,   7.9 &  11.2 &   89.8 &17$^{\rm h}$34$^{\rm m}$34$^{\rm s}\!.$0219 & -33$^{\circ}$33$^{\prime}$08$^{\prime \prime}\!\!.$2368 &   91.5 &  110.7  \\
 19 & 111.6 ,   7.9 &  14.1 &  127.8 &17$^{\rm h}$34$^{\rm m}$19$^{\rm s}\!.$3574 & -33$^{\circ}$31$^{\prime}$53$^{\prime \prime}\!\!.$7939 &   81.2 &   76.5  \\
 20 & 149.2 ,   8.1 &  18.3 &   95.9 &17$^{\rm h}$34$^{\rm m}$23$^{\rm s}\!.$7535 & -33$^{\circ}$31$^{\prime}$52$^{\prime \prime}\!\!.$5611 &   55.4 &   78.1  \\
 21 & 323.2 ,   8.3 &  39.1 &  316.6 &17$^{\rm h}$34$^{\rm m}$50$^{\rm s}\!.$7384 & -33$^{\circ}$31$^{\prime}$33$^{\prime \prime}\!\!.$1535 &   26.3 &   27.8  \\
 22 &  56.4 ,   7.2 &   7.8 &  119.5 &17$^{\rm h}$34$^{\rm m}$31$^{\rm s}\!.$0282 & -33$^{\circ}$31$^{\prime}$32$^{\prime \prime}\!\!.$1939 &  100.3 &  104.0  \\
 25 &  73.0 ,   7.7 &   9.4 &  457.4 &17$^{\rm h}$33$^{\rm m}$51$^{\rm s}\!.$5883 & -33$^{\circ}$31$^{\prime}$20$^{\prime \prime}\!\!.$5649 &  163.2 &  104.0  \\
 30 & 120.8 ,   9.1 &  13.3 &  648.8 &17$^{\rm h}$33$^{\rm m}$38$^{\rm s}\!.$8158 & -33$^{\circ}$29$^{\prime}$16$^{\prime \prime}\!\!.$3236 &  106.3 &   79.9  \\
 31 & 461.4 ,   8.1 &  56.9 &  262.9 &17$^{\rm h}$34$^{\rm m}$32$^{\rm s}\!.$8634 & -33$^{\circ}$29$^{\prime}$07$^{\prime \prime}\!\!.$8686 &   18.9 &   19.1  \\
 33 & 107.0 ,   8.3 &  12.9 &  337.7 &17$^{\rm h}$34$^{\rm m}$38$^{\rm s}\!.$3613 & -33$^{\circ}$28$^{\prime}$14$^{\prime \prime}\!\!.$2692 &   83.5 &   75.8  \\
 35 & 134.5 ,   8.6 &  15.6 &  491.0 &17$^{\rm h}$34$^{\rm m}$23$^{\rm s}\!.$9488 & -33$^{\circ}$25$^{\prime}$10$^{\prime \prime}\!\!.$4081 &   76.0 &   80.2  \\
\hline
\ \end{tabular}
\label{tab:fit_params}
\tablefoot{
  \tablefoottext{p}{This is PSR J1734$-$3333.}
  }
\end{table*}

\begin{table*}
\centering
\caption{Source parameters from the 2019 map obtained with the CASA routine {\sc imfit}.}
\label{tab:src2}
\begin{tabular}{lHKKccKK}
\hline
\hline
\multicolumn{1}{c}{ID}
& \multicolumn{1}{c}{Flux Density}  
& \multicolumn{1}{c}{S/N}
& \multicolumn{1}{c}{Dist. PSR}
& \multicolumn{1}{c}{RA}  
& \multicolumn{1}{c}{DEC}
& \multicolumn{1}{c}{1-$\sigma$ RA uncertainty}  
& \multicolumn{1}{c}{1-$\sigma$ DEC uncertainty } \\ 
\multicolumn{1}{c}{ }
& \multicolumn{1}{c}{($\mu{\rm Jy}$)}  
& \multicolumn{1}{c}{ }
& \multicolumn{1}{c}{(arcsec)}
& \multicolumn{1}{c}{J2000}  
& \multicolumn{1}{c}{J2000}
& \multicolumn{1}{c}{(mas) }  
& \multicolumn{1}{c}{(mas) }\\
\hline
 3 &  71.4 ,   8.2 &   8.7 &  554.5 &17$^{\rm h}$34$^{\rm m}$31$^{\rm s}\!.$0092 & -33$^{\circ}$42$^{\prime}$32$^{\prime \prime}\!\!.$1128 &   25.5 &  128.6  \\
 4 & 121.0 ,   8.1 &  15.0 &  388.8 &17$^{\rm h}$34$^{\rm m}$19$^{\rm s}\!.$2163 & -33$^{\circ}$39$^{\prime}$36$^{\prime \prime}\!\!.$7490 &   14.2 &   85.8  \\
 5 & 202.2 ,   7.8 &  26.0 &  349.6 &17$^{\rm h}$34$^{\rm m}$33$^{\rm s}\!.$1663 & -33$^{\circ}$39$^{\prime}$00$^{\prime \prime}\!\!.$7171 &    8.8 &   47.9  \\
 7 &  68.0 ,   9.6 &   7.1 &  706.0 &17$^{\rm h}$33$^{\rm m}$36$^{\rm s}\!.$7158 & -33$^{\circ}$38$^{\prime}$44$^{\prime \prime}\!\!.$4313 &   58.3 &  143.1  \\
 8 & 104.2 ,   8.8 &  11.9 &  364.9 &17$^{\rm h}$34$^{\rm m}$04$^{\rm s}\!.$9742 & -33$^{\circ}$37$^{\prime}$21$^{\prime \prime}\!\!.$0350 &   21.3 &   98.7  \\
 9 & 212.0 ,   9.3 &  22.8 &  533.5 &17$^{\rm h}$35$^{\rm m}$05$^{\rm s}\!.$4935 & -33$^{\circ}$37$^{\prime}$08$^{\prime \prime}\!\!.$1699 &   12.2 &   45.4  \\
 10 &  43.4 ,   7.7 &   5.7 &  273.3 &17$^{\rm h}$34$^{\rm m}$41$^{\rm s}\!.$7537 & -33$^{\circ}$36$^{\prime}$40$^{\prime \prime}\!\!.$6089 &   46.7 &  161.0  \\
 11 &  79.5 ,   8.2 &   9.7 &  196.0 &17$^{\rm h}$34$^{\rm m}$29$^{\rm s}\!.$3628 & -33$^{\circ}$36$^{\prime}$33$^{\prime \prime}\!\!.$5833 &   26.3 &   95.1  \\
 12 &  61.5 ,   9.5 &   6.5 &  568.7 &17$^{\rm h}$35$^{\rm m}$09$^{\rm s}\!.$9587 & -33$^{\circ}$36$^{\prime}$23$^{\prime \prime}\!\!.$9360 &   52.9 &  150.8  \\
 13 &  42.9 ,   9.0 &   4.8 &  226.8 &17$^{\rm h}$34$^{\rm m}$38$^{\rm s}\!.$9721 & -33$^{\circ}$36$^{\prime}$09$^{\prime \prime}\!\!.$2950 &   49.8 &  163.9  \\
 14 & 115.2 ,   8.8 &  13.1 &  340.0 &17$^{\rm h}$34$^{\rm m}$00$^{\rm s}\!.$2709 & -33$^{\circ}$34$^{\prime}$29$^{\prime \prime}\!\!.$4341 &   18.6 &   77.8  \\
 17$^{p}$ & 299.9 ,   7.6 &  39.4 &    2.8 &17$^{\rm h}$34$^{\rm m}$27$^{\rm s}\!.$0561 & -33$^{\circ}$33$^{\prime}$21$^{\prime \prime}\!\!.$9422 &    6.1 &   31.0  \\
 18 &  81.5 ,   7.8 &  10.5 &   90.0 &17$^{\rm h}$34$^{\rm m}$34$^{\rm s}\!.$0394 & -33$^{\circ}$33$^{\prime}$08$^{\prime \prime}\!\!.$5021 &   19.6 &  108.9  \\
 19 &  83.1 ,   8.5 &   9.8 &  128.0 &17$^{\rm h}$34$^{\rm m}$19$^{\rm s}\!.$3521 & -33$^{\circ}$31$^{\prime}$53$^{\prime \prime}\!\!.$5162 &   32.9 &  110.5  \\
 20 & 137.5 ,   7.4 &  18.6 &   95.9 &17$^{\rm h}$34$^{\rm m}$23$^{\rm s}\!.$7530 & -33$^{\circ}$31$^{\prime}$52$^{\prime \prime}\!\!.$5326 &   12.7 &   66.4  \\
 21 & 248.9 ,   8.5 &  29.1 &  316.6 &17$^{\rm h}$34$^{\rm m}$50$^{\rm s}\!.$7281 & -33$^{\circ}$31$^{\prime}$32$^{\prime \prime}\!\!.$9706 &    9.7 &   35.0  \\
 22 &  35.9 ,   7.4 &   4.8 &  119.7 &17$^{\rm h}$34$^{\rm m}$31$^{\rm s}\!.$0308 & -33$^{\circ}$31$^{\prime}$32$^{\prime \prime}\!\!.$0615 &   91.8 &  251.3  \\
 25 &  44.7 ,   9.7 &   4.6 &  457.5 &17$^{\rm h}$33$^{\rm m}$51$^{\rm s}\!.$5749 & -33$^{\circ}$31$^{\prime}$20$^{\prime \prime}\!\!.$5771 &   70.9 &  186.3  \\
 30 &  42.8 ,   8.8 &   4.8 &  648.9 &17$^{\rm h}$33$^{\rm m}$38$^{\rm s}\!.$8149 & -33$^{\circ}$29$^{\prime}$16$^{\prime \prime}\!\!.$0670 &   82.9 &  288.1  \\
 31 & 384.0 ,   7.7 &  49.9 &  262.9 &17$^{\rm h}$34$^{\rm m}$32$^{\rm s}\!.$8613 & -33$^{\circ}$29$^{\prime}$07$^{\prime \prime}\!\!.$8548 &    4.6 &   23.6  \\
 33 &  58.0 ,   7.2 &   8.0 &  337.7 &17$^{\rm h}$34$^{\rm m}$38$^{\rm s}\!.$3459 & -33$^{\circ}$28$^{\prime}$14$^{\prime \prime}\!\!.$1778 &   50.5 &  212.3  \\
 35 & 122.9 ,   8.1 &  15.1 &  491.0 &17$^{\rm h}$34$^{\rm m}$23$^{\rm s}\!.$9361 & -33$^{\circ}$25$^{\prime}$10$^{\prime \prime}\!\!.$4301 &   19.3 &   73.5  \\
\hline
\ \end{tabular}
\label{tab:fit_params}
\tablefoot{
  \tablefoottext{p}{This is PSR J1734$-$3333.}
  }
\end{table*}

After calibration, we used the {\sc tclean} task from CASA
(version 5.8) to image the measurement sets for each epoch. We tested
several weighting and imaging parameters and we settled for the ones
that provided better sensitivity and lower level of imaging
artefacts. We chose the {\it Briggs} weighting scheme \citepads{1999ASPC..180..127B} with {\sc robust=0.5} in order to find a balance between good signal-to-noise
ratio and a narrow synthesised beam.  Also, we used the option {\sc
  nterms}=2, which allows the {\sc tclean} task to model the spectral
indices of the sources. We created images using a large grid of
$13122\times13122$ square pixels, each with a side of
0.1\,arcsec. This image size is roughly 1.5 times the primary beam at
S-band, motivated by the intention of also including in the analysis sources
that lie farther away from the position of the pulsar. Because of the
large image size, we had to use the wide-field imaging parameters
{\mbox{\sc gridder}}={\em wproject} and {\sc wprojectplanes}=-1.

\begin{figure}
	\centering
   	\includegraphics[width=4.4cm]{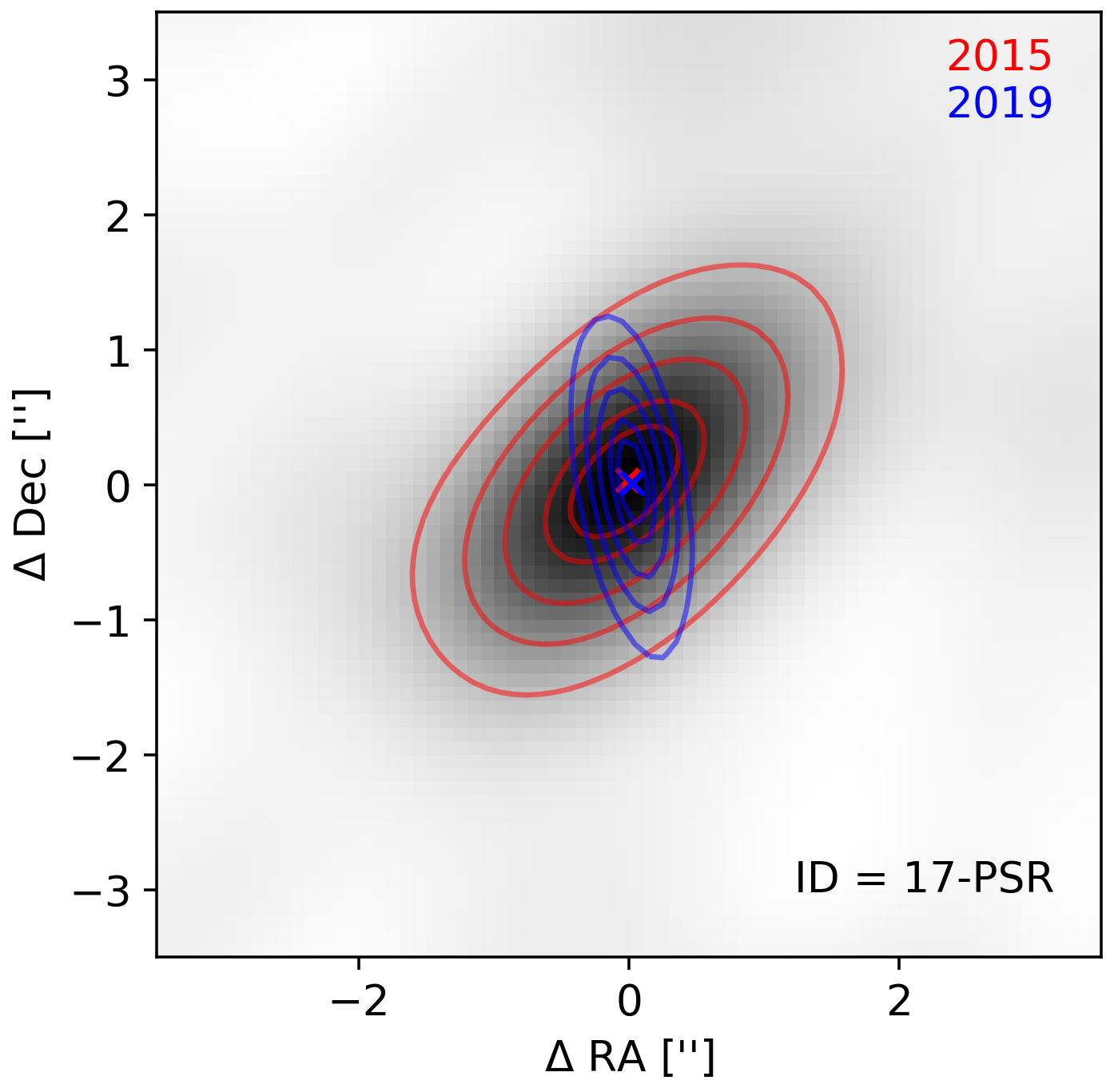}
	\includegraphics[width=4.4cm]{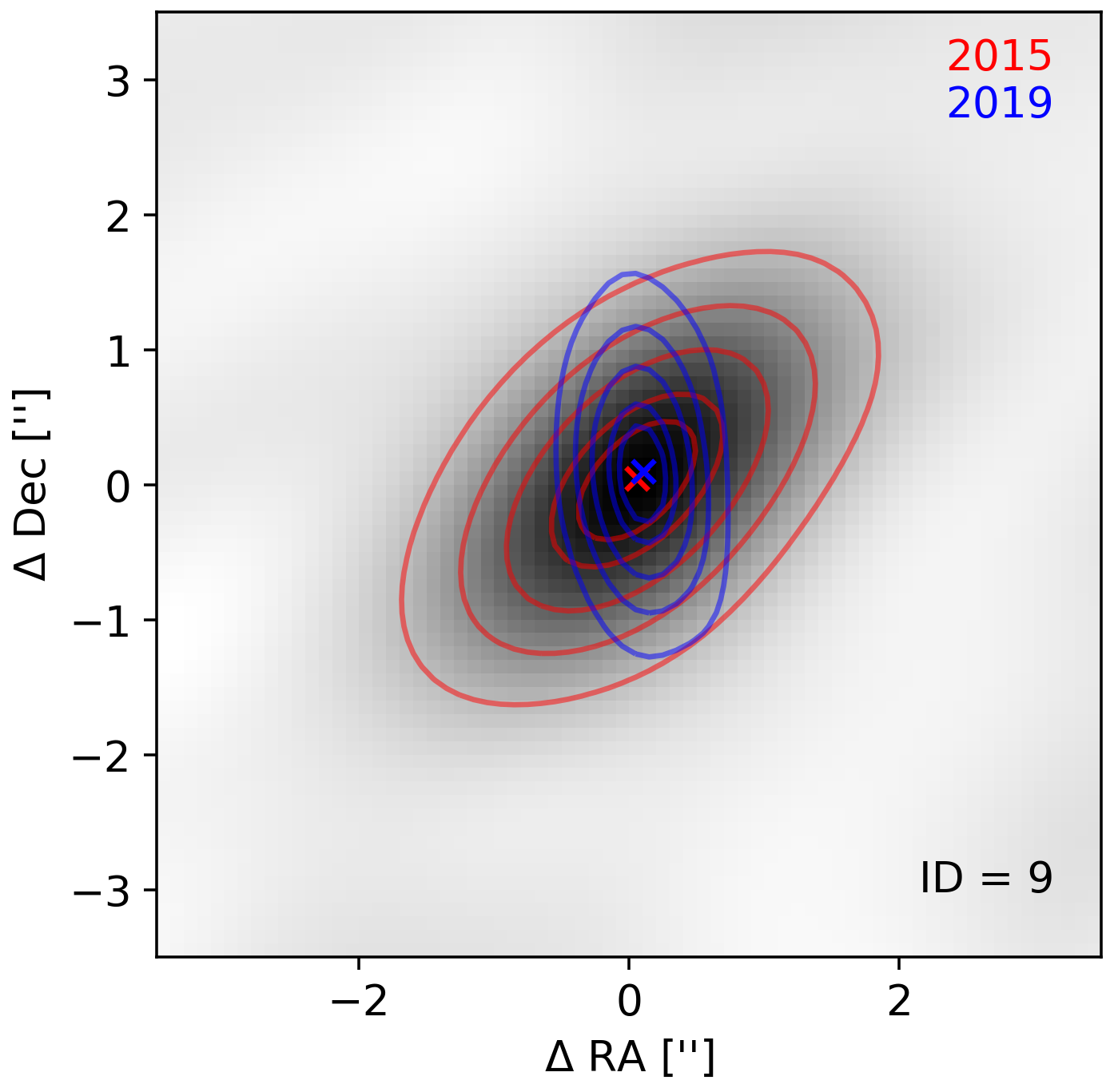}  
	   	\includegraphics[width=4.4cm]{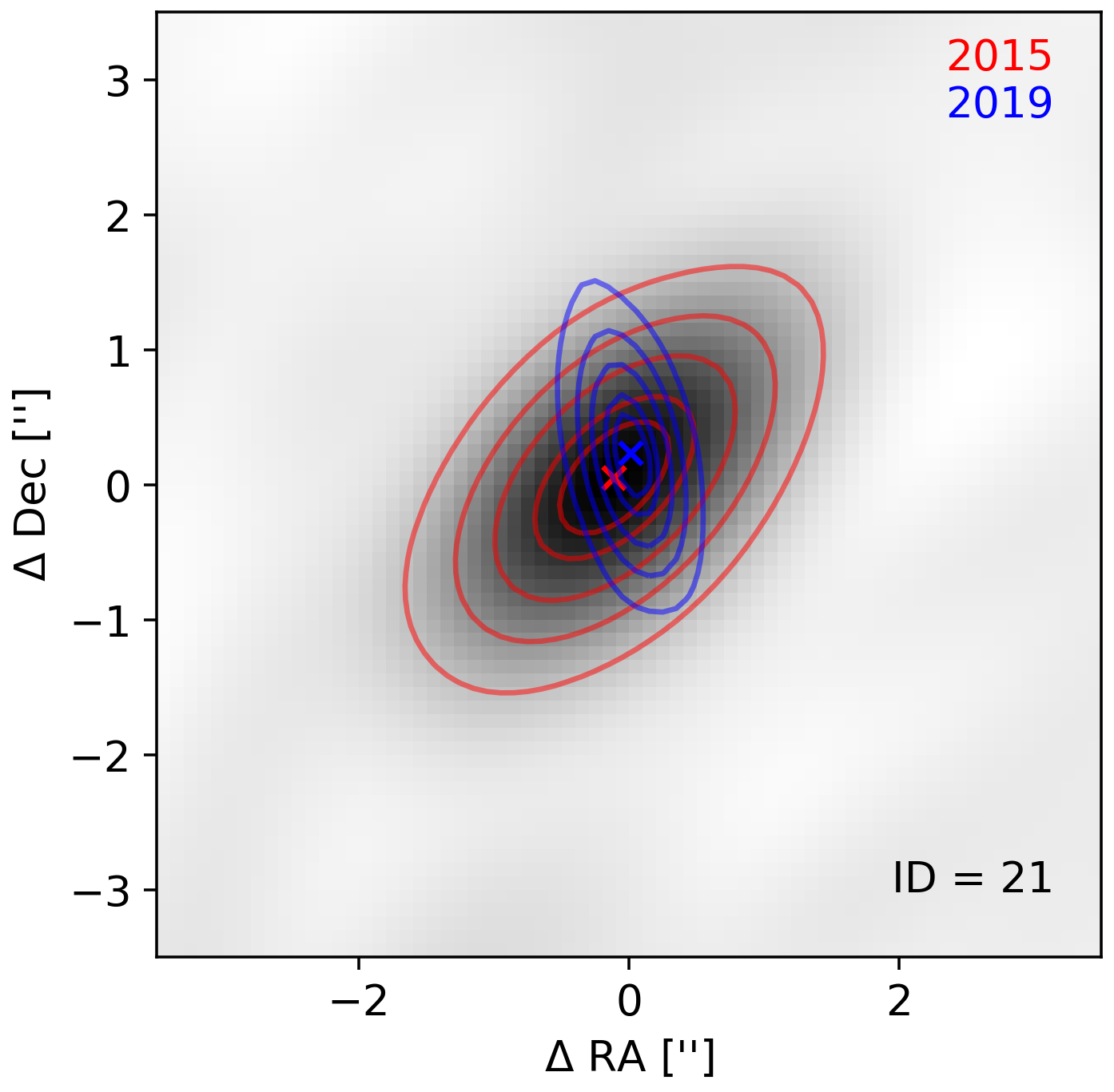}
   	\includegraphics[width=4.4cm]{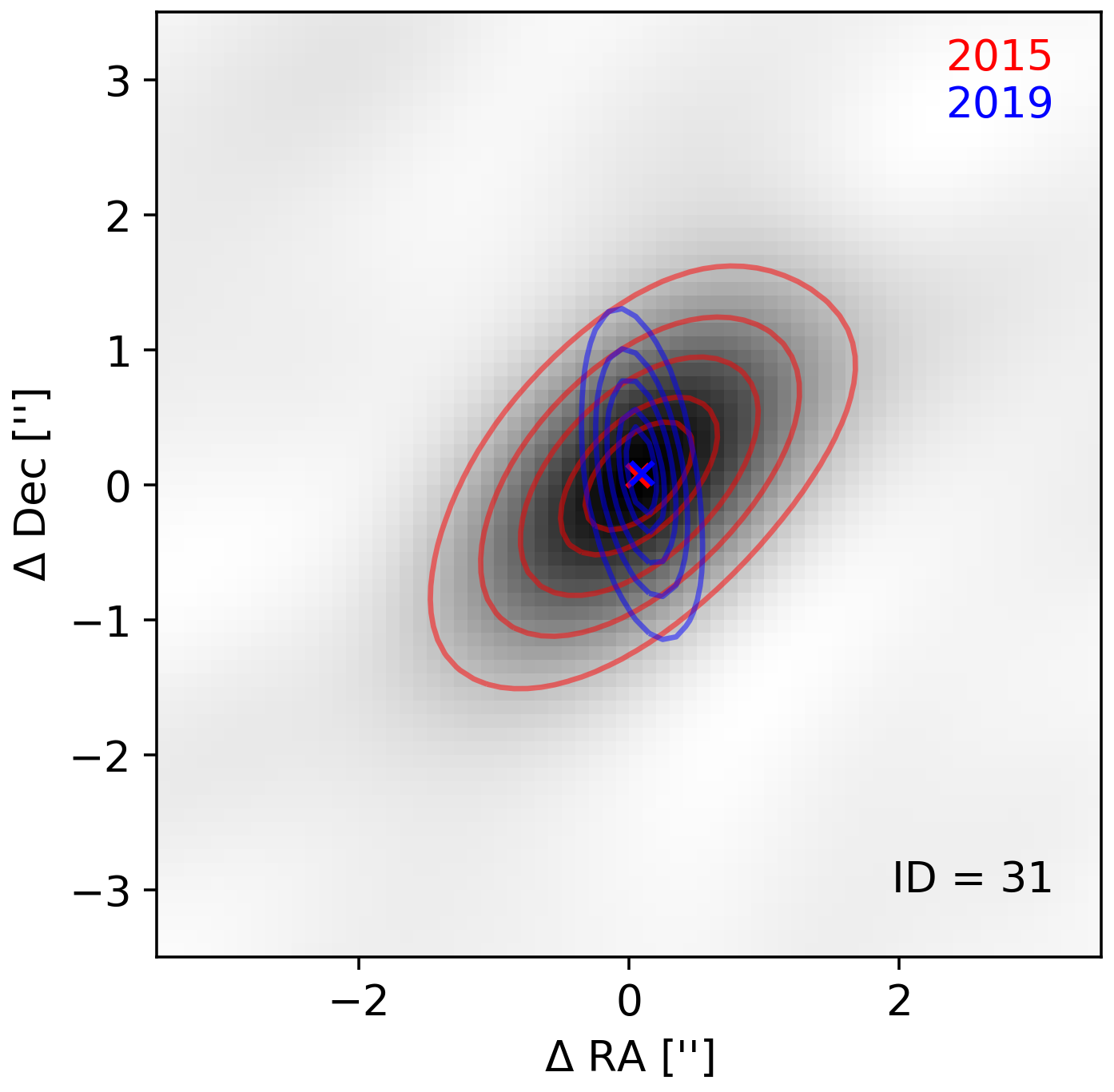}
    \caption{Clean S-band images for the four brighter sources in the 2015 observation. Overlaid contours for the 2015 (red) and the 2019 (blue) observations are shown too. The pulsar is source 17 (top left).
}
    \label{fig:maps}
\end{figure}

Due to the different array configurations for each epoch, the 2019
data have a more elongated synthesised beam (see Table \ref{tab:log}). 
Figure \ref{fig:maps} shows the 2015 S-band image for the pulsar and for the three brighter sources in the field. Contours of the 2019 S-band observation are overlaid. 

\subsection{Source detection and astrometry}
\label{sec:sources}
After we produced cleaned images for each epoch, we used an automatic source detection software (the {\sc DAOStarFinder} routine from the {\sc photutils} package\footnote{\url{https://photutils.readthedocs.io/en/stable/index.html}}) to identify all the sources that could be used to reference the position of the pulsar. This procedure yielded a total of 41 sources that were detected in both epochs. From these sources, a number of them were complex, i.e. they were resolved in the 2019 data but unresolved in the 2015 maps (due to the difference in beam sizes) while others were resolved and showed extended emission in both epochs. We kept only the ones that were consistent with being unresolved point sources in both epochs. The final number of sources that we used to measure their positions was $N=\N$.

We fitted for the position of the selected sources using the {\sc imfit} task from CASA. 
The task fits a 2-D Gaussian for each source where the free parameters are the location of the centroid, both semi-axis and its position angle. In tables \ref{tab:src1} and \ref{tab:src2} we list the parameters for the sources, including the pulsar, for both epochs. In these tables, the ID number is the index of the source from the original 41 sources detected by {\sc DAOStarFinder}. We also list the flux density of each source, its signal-to-noise ratio (S/N) based on the rms noise of the maps, the distance to the coordinates of the pulsar, as previously published in \citetads[][RA(J2000)=17h34m26.9s,   DEC(J2000)=-33d33m20s]{2011ApJ...741L..13E} 
and the RA-DEC coordinates with their respective 1-$\sigma$ uncertainty as calculated by {\sc imfit}. 
The listed flux densities have not been corrected by the primary beam attenuation. In these tables, the pulsar corresponds to  ID=17.

\section{Proper motion estimate}
At each epoch, we measure the position of the pulsar and the point-like background sources listed in Tables \ref{tab:src1} and \ref{tab:src2}. 
We assume that the proper motion of all these background sources is zero, and use that constraint to measure a single offset, $\Delta C$, between the two observations.
The ensemble of background sources (or reference sources) is characterised via its centroid, which is the average of their positions at a given epoch: $C=\left(\langle\alpha\rangle, \langle\delta\rangle\right)$. 
Thus the offset of the centroid is $\Delta C=C_{2019}-C_{2015}$.
This quantity is used to minimise the net translation that this ensemble of sources may present between the two observations.
The effective displacement of the pulsar between the two epochs is the position difference between the 2019 and 2015 observations, $\Delta p$, corrected by $\Delta C$: 
\begin{equation}
  D=\Delta p - \Delta C \quad.
\end{equation}

The uncertainty of $D$ was calculated by propagating the uncertainties
of $\Delta p$ and $\Delta C$.  To calculate the uncertainty of $\Delta
p$, the error bars reported by {\sc imfit} (Tables \ref{tab:src1} and
\ref{tab:src2}) were added in quadrature.  
The uncertainty of $\Delta C$ was calculated by adding the uncertainty of $C_{2015}$ and $C_{2019}$ in quadrature, where each of them was measured using a bootstrap method \citepads[][]{1991Sci...253..390E}. 
In this method, several realisations (500,000 in this case) of the calculation of $\Delta C$ were done using samples of reference sources which were variations of the original sample of $N$ sources (in this case the $N=\N$ sources in Tables \ref{tab:src1} and \ref{tab:src2}).  
Each varied sample was generated by selecting $N$ reference sources randomly from the original list of $N$ sources; and note that any source could be selected more than once in a given realisation.  
The standard deviation of all the offsets was defined as the uncertainty of the centroid offset calculated from the original sample.

\begin{figure}
	\centering
   	\includegraphics[width=9cm]{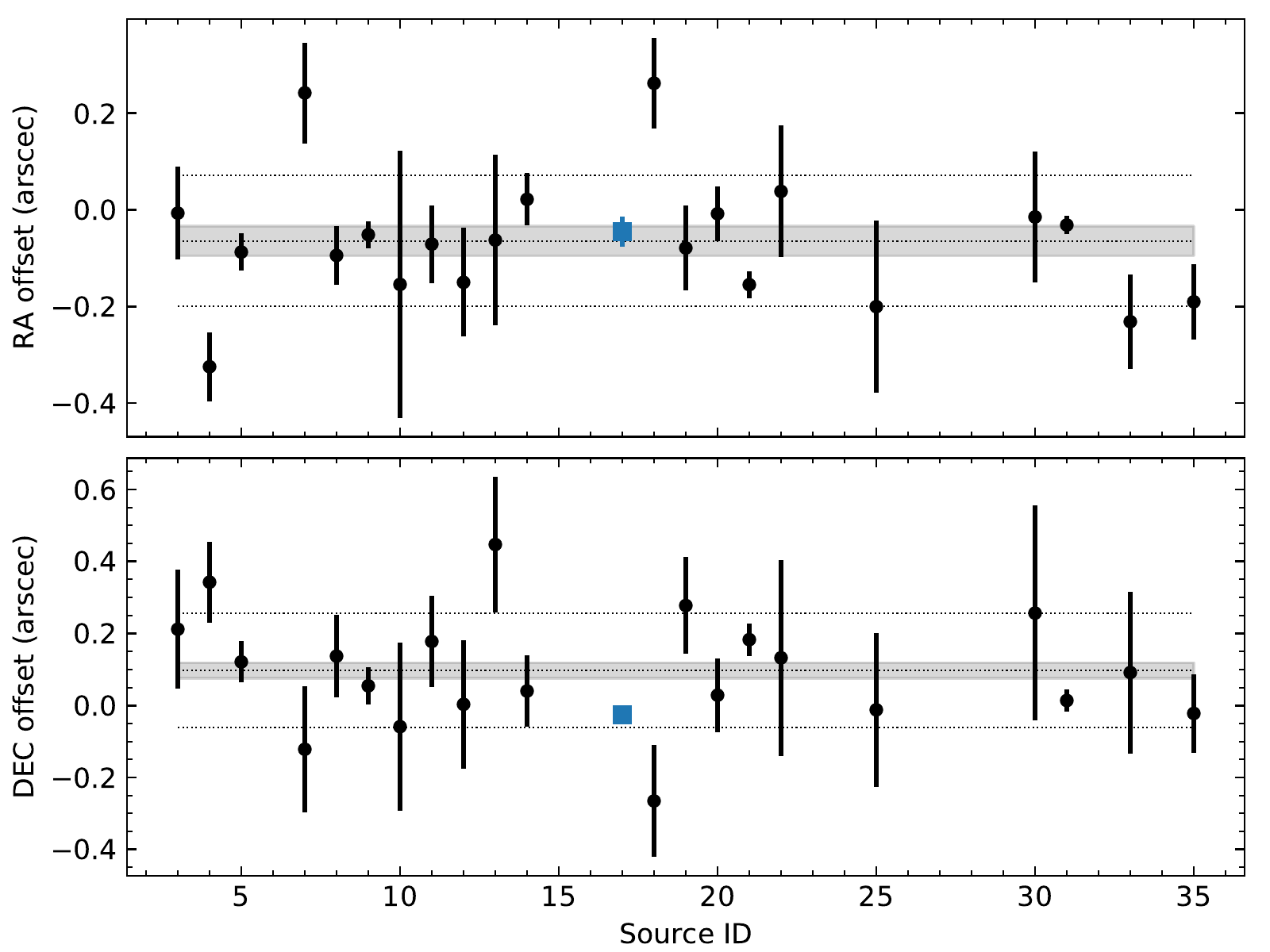}
   	\includegraphics[width=9cm]{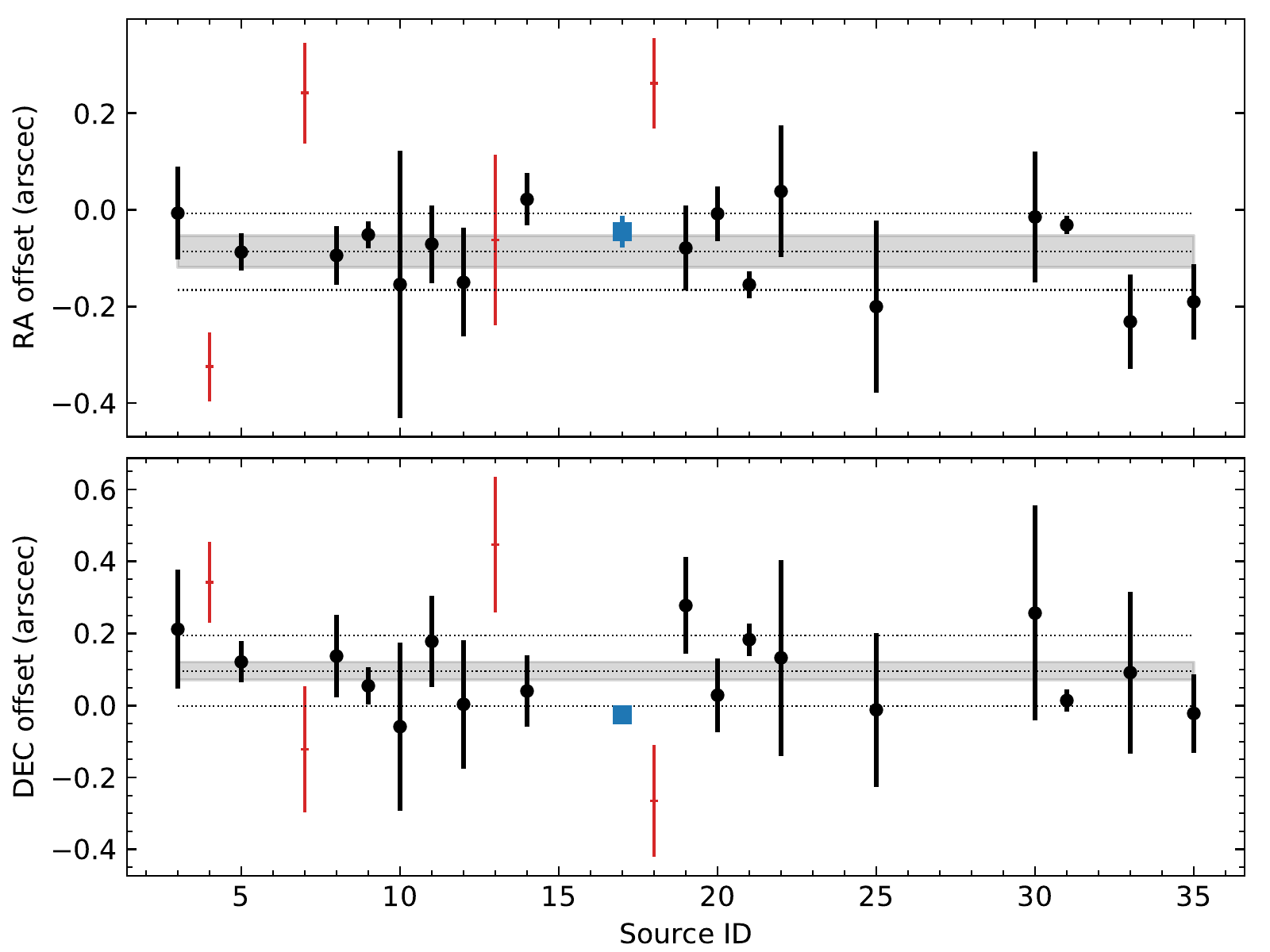}
    \caption{Offsets between the positions measured in 2015 and 2019 of the sources in Tables \ref{tab:src1} and \ref{tab:src2}.
    The pulsar is source 17 and its offset ($\Delta p$) is plotted with a blue square.
    The centroid's offset and the standard deviation of the reference sources around this value are shown with horizontal dotted lines.
    The grey area in each plot is the uncertainty of the centroid's offset, as calculated using a bootstrap method.
    {\it Top}: Results for RA and DEC obtained for \N\ reference sources. 
    {\it Bottom}: Results obtained for the \Nb\ reference sources that present offsets  consistent with the centroid's offset.
    The rejected sources are plotted with smaller red symbols and were not used to measure the centroid (see Table \ref{tab:disps}).
    }
    \label{fig:offsets}
\end{figure}

Using the \N\ reference sources in Tables \ref{tab:src1} and \ref{tab:src2} we obtained a displacement for the pulsar in RA of $0\farcs02\pm0\farcs04$, and a displacement in DEC of
$-0\farcs12\pm0\farcs05$ (Table \ref{tab:disps}).
Figure \ref{fig:offsets} plots the offset in RA and DEC for the sources listed in Tables \ref{tab:src1} and \ref{tab:src2}. 
The offsets between epochs of some individual sources can be significantly different to the centroid's offset, and sometimes even larger than the offset of the pulsar.  
For example, source 18 has particularly large residuals, as can be seen in the top panel of Fig. \ref{fig:offsets}, probably due to its own proper motion. 
In order to test the effect of this large shift in position of individual sources, we experimented by removing the reference sources with large residuals and calculated everything again. 
The results obtained for most combinations were consistent (considering uncertainties) with the results described above: a marginal detection of the pulsar displacement in DEC (of negative sign) and a non-detection of the pulsar displacement in RA. 
As an example, in Table \ref{tab:disps} and Fig. \ref{fig:offsets} (bottom) we present the results obtained when using only the \Nb\ sources with offsets that lie within the standard deviation of the entire sample (that is, are between the dotted lines in the top plots in Fig. \ref{fig:offsets}). 
The difference in the resulting proper motion estimate when using both samples is consistent with zero. 
These results indicate that both the reference frame and the measured proper motion for the pulsar are robust.

\begin{table}
  \centering
  \caption{Offsets and displacements for two sets of reference sources.}
  \label{tab:disps}
  \begin{tabular}{lHH}
    \hline \hline
    \multicolumn{1}{l}{Parameter} &
    \multicolumn{1}{c}{RA} &
    \multicolumn{1}{c}{DEC} \\
    \hline
    $\Delta p$ (\arcsec) &   -0.04 , 0.03  &  -0.03 , 0.04  \\
    \hline
    \multicolumn{3}{l}{\sl Using all the reference sources} \\
    $\Delta C$ (\arcsec) &   -0.06 , 0.03  &  0.10 , 0.02  \\
    $D$        (\arcsec) &    0.02 , 0.04  & -0.12 , 0.05  \\
    Proper motion (mas\,yr$^{-1}$)&     5 , 10    & -29   , 11    \\
    \hline
    \multicolumn{3}{l}{\sl Using \Nb\ reference sources$^{(a)}$} \\
    $\Delta C$ (\arcsec) &   -0.09 , 0.03  &  0.10 , 0.02  \\
    $D$        (\arcsec) &    0.04 , 0.04  & -0.12 , 0.05  \\
    Proper motion (mas\,yr$^{-1}$)&     10 , 10    & -29   , 11    \\
    \hline
  \end{tabular}
  \tablefoot{The offset of the pulsar is $\Delta p$. 
  The offset of the reference frame ($\Delta C$) and pulsar effective displacement ($D$) are given for two sets of reference sources.\\
    \tablefoottext{a}{Having rejected the sources with ID 4, 7, 13, and 18. 
  }}
\end{table}

\begin{figure}
	\centering
   	\includegraphics[width=9cm]{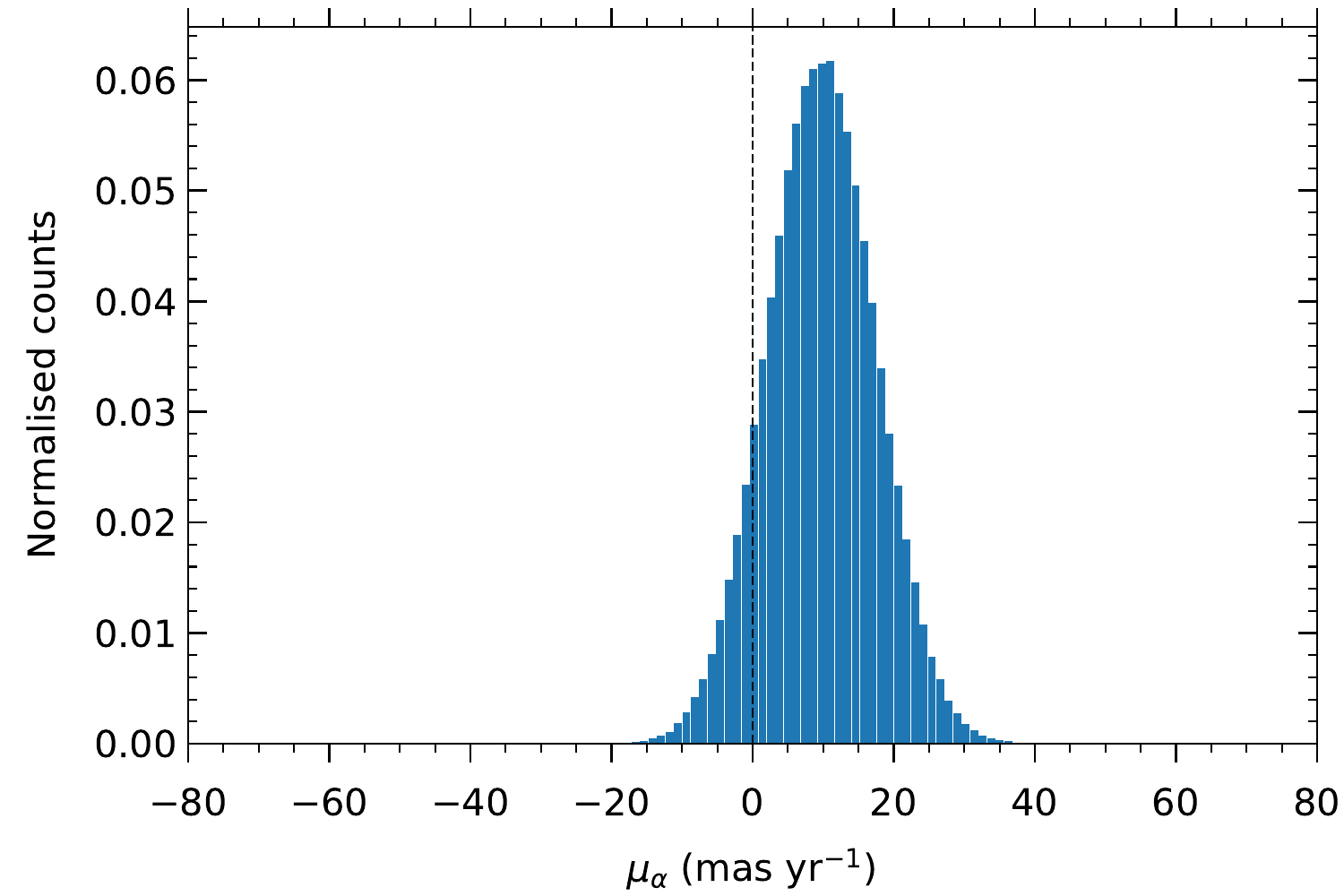}
   	\includegraphics[width=9cm]{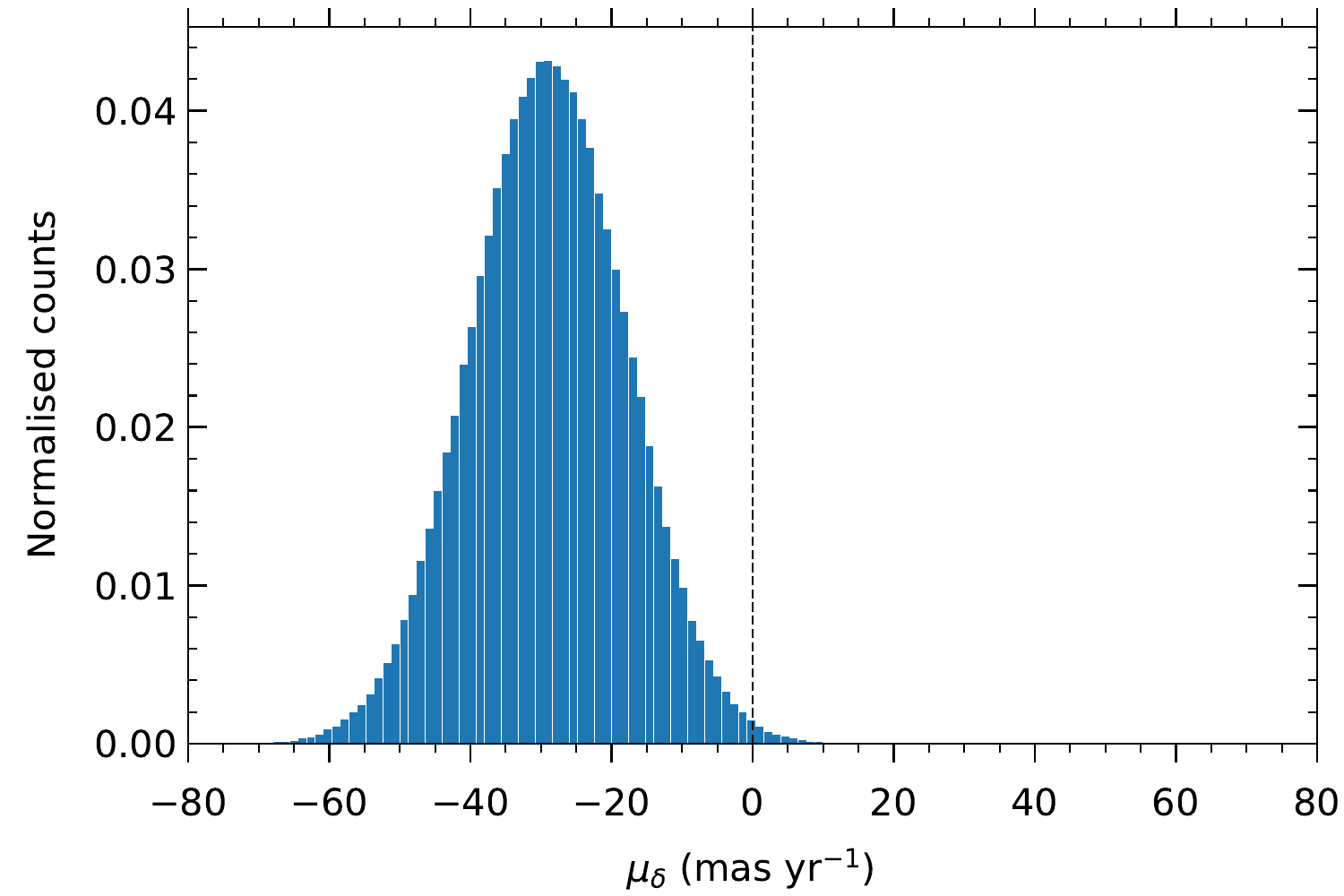}
    \caption{Pulsar proper motion values obtained from realisations of both the centroid of 17 reference sources (from the bootstrap method) and the pulsar offset $\Delta p$ (using a Gaussian distribution centred at the measured value and with a dispersion given by its uncertainty, Table \ref{tab:disps}).
    }
    \label{fig:motions}
\end{figure}

Distributions of the proper motion components $\mu_\alpha$ and $\mu_\delta$, for the analysis that uses 17 reference sources, are shown in Fig. \ref{fig:motions}. 
Each value comes from a measurement of $D$ performed via a particular realisation of the frame of reference (using the bootstrap process described above) and the pulsar offset $\Delta p$.
For $\Delta p$ we drew values from normal distributions centred at the RA and DEC offset components presented in Table \ref{tab:disps} and with dispersions equal to the quoted uncertainties.
Note that the distribution for $\mu_\delta$ is wider mainly because of the orientation of the elongated beam shape in the 2019 observation, which covers a longer range in DEC than in RA (Fig. \ref{fig:maps}).
The proper motions were calculated for a difference in time of $4.2433$\,yr between observations.
Our results are consistent with a motion of the pulsar between both epochs only in DEC, which implies a proper motion $\mu_\delta=-29\pm11$\,mas\,yr$^{-1}$. 
We acknowledge that this is a marginal detection, as the displacement of the pulsar
in DEC is similar to the standard deviation of the other sources displacements (the horizontal dotted lines in Fig. \ref{fig:offsets}).

\section{A complicated association with G354.8$-$0.8}
If PSR J1734$-$3333 and SNR G354.8$-$0.8 were associated, the proper motion of the pulsar should be $-172$\,mas\,yr$^{-1}$ in RA and $64$\,mas\,yr$^{-1}$ in DEC, assuming an age of $8.1$\,kyr (the characteristic age of the pulsar) and that it has travelled with constant velocity since the explosion.  
From our results, we can write the $2$-$\sigma$ limits $-10<\mu_\alpha<30$\,mas\,yr$^{-1}$ (Fig. \ref{fig:motions}, Table \ref{tab:disps}), which is inconsistent with the expected $-172$\,mas\,yr$^{-1}$ at high significance.
In declination, our detection $\mu_\delta=-29\pm11$\,mas\,yr$^{-1}$ is also significantly inconsistent with the expectation, and implies movement of the pulsar but towards the SNR. 
These results therefore indicate that the pulsar and G354.8--0.8 are not related, meaning that the original SNR that was born with the pulsar might have faded and the system may be older than $10$-$100$\,kyr.

We note that $\mu_\delta=-29\pm11$\,mas\,yr$^{-1}$ implies a transverse velocity of $630\pm210$\,km\,s$^{-1}$ \citepads[at a distance $d_{4.5}=4.5$\,kpc, as estimated from the dispersion measure and the YMW16 electron density model,][]{2017ApJ...835...29Y}, which is faster than the average transverse speed amongst pulsars but still a possible value.
The distance to the pulsar is however rather uncertain, as it depends on the model used for the
electron density column.  
The NE2001 model \citepads{2002astro.ph..7156C} indicates that the distance is $d_{6.1}=6.1$\,kpc. 
This distance would imply a speed of $870\pm290$\,km\,s$^{-1}$, which lies at the very high end of the distribution of pulsar transverse velocities \citepads{2005MNRAS.360..974H}.

However, the possibility that the marginal detection of $\mu_\delta$ is a statistical fluke cannot be completely neglected. 
Hence we also consider the possibility that the proper motion of the pulsar is below our detection capabilities, and consider a $2$-$\sigma$ upper limit for the total proper motion $\mu<28$\,mas\,yr$^{-1}$, obtained by doubling the calculated uncertainties in RA and DEC.  
Given this proper motion constraint, if pulsar and SNR were indeed associated, the pulsar would need at least $45$\,kyr to cover the distance determined by the $21$\arcmin\ offset with the SNR centre.
If the system was older than $45$\,kyr, as required by this upper limit scenario 
and for a true SNR-pulsar association, the pulsar transverse velocity would be less than $590\, d/d_{4.5}$\,km\,s$^{-1}$, where $d$ is the assumed distance to the pulsar.  
This is however an unlikely scenario given the results of our observations, which suggest that the pulsar moves in a direction $130^\circ$ away of what the association with the SNR requires (with a measured position angle error of $16^\circ$, this is a $8$-$\sigma$ distance in position angle).

\subsection{Another SNR as candidate?}
\begin{figure}
  \centering
  \includegraphics[width=9cm]{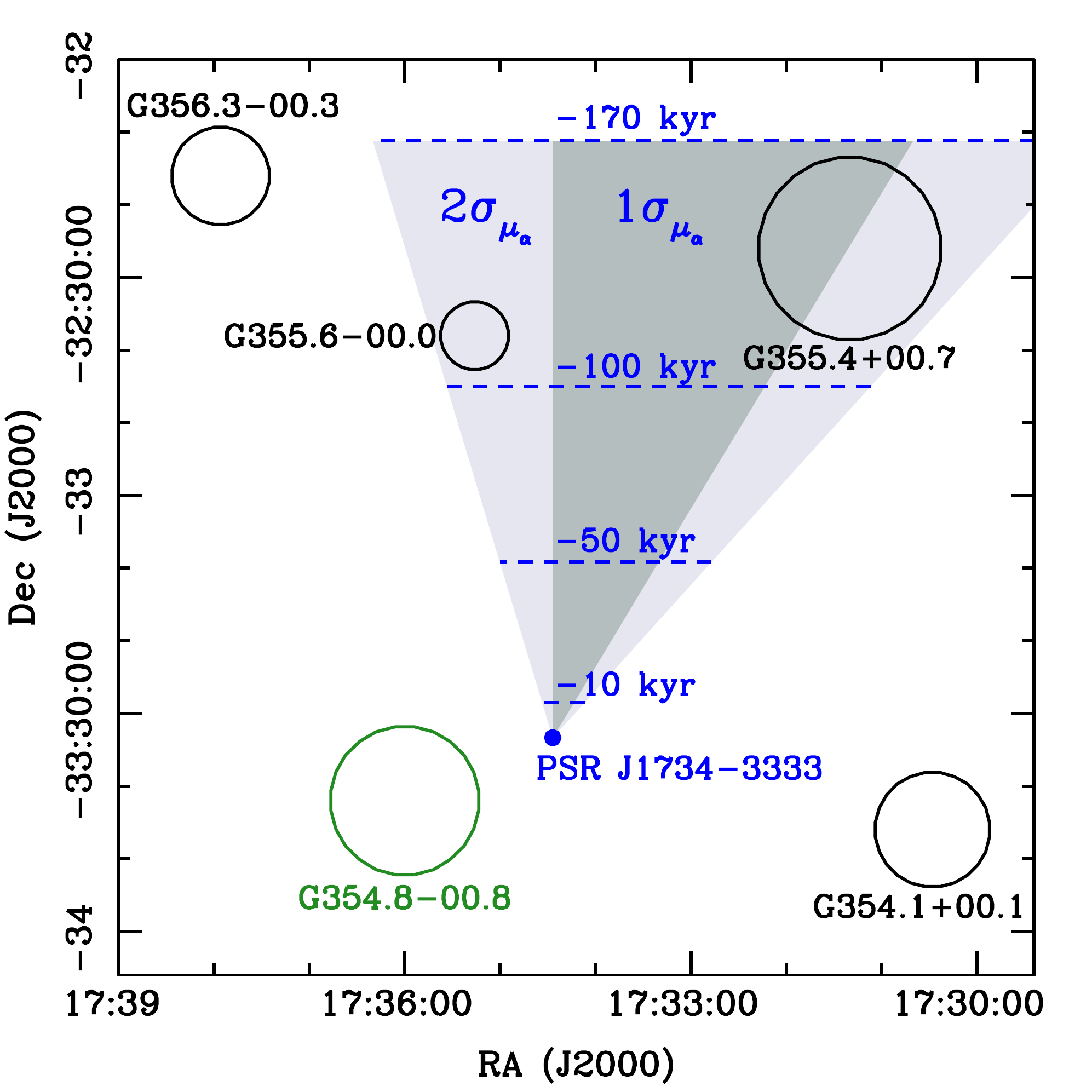}
  \caption{Supernova remnants in the vicinity of PSR\,J1734$-$3333. 
    The black circles represent SNRs from the \citetads{2019JApA...40...36G} catalogue and SNR G354.8$-$0.8 is in green. 
    The diameter of each circle is given by the size of the SNR listed in the catalogue. 
    The blue dot shows the current location of the pulsar, and the horizontal dashed lines show where the pulsar would have been at each labeled time in the past considering a proper motion $\mu_\delta=-29$\,mas\,yr$^{-1}$.
    The shaded regions mark the possible location of the pulsar in the past considering also a proper motion in RA, for which we use the $1$-$\sigma$ range $\mu_\alpha\in[0, 20]$\,mas\,yr$^{-1}$ (dark shading), and the  $2$-$\sigma$ range $\mu_\alpha\in[-10, 30]$\,mas\,yr$^{-1}$ (light shading). 
    }
  \label{fig:snrs}
\end{figure}

Figure \ref{fig:snrs} shows the known SNRs that lie in the vicinity of PSR\,J1734$-$3333. 
Their positions were taken from the \citetads{2019JApA...40...36G} catalogue of supernova remnants. 
We also searched for SNRs detected at lower radio wavelengths, hence potentially older than those detected at shorter wavelengths \citepads{2006JPhCS..54..152R,2019PASA...36...45H}, but found none near the pulsar.
In Fig. \ref{fig:snrs}, regions that show where the pulsar was in the past are plotted, based on the proper motion we obtained.
Just by considering the direction of the proper motion, it is tempting to consider either G355.6$-$00.0 or G355.4+00.7 as candidates to the associated SNR for the pulsar.  
In terms of the direction, this is plausible given that $\mu_{\alpha}$ is essentially unconstrained by our data. 
However, the large separation between the PSR and both SNRs makes the association unlikely, as it would require an age for the SNRs greater than 100\,kyr, and it is unlikely that SNRs remain detectable in the radio for this long \citepads{2017MNRAS.464.2326S}.   
Moreover, G355.6+00.0 is believed to lie at a distance of around 12\,kpc, albeit with large uncertainties \citepads{1998ApJ...504..761C,2013PASJ...65...99M}, which is much larger than what it is estimated for the pulsar. 
For this remnant, the age is estimated to be $\sim20$\,kyr by \citetads{2013PASJ...65...99M} and $9\pm1.7$\,kyr by \citetads{2020ApJS..248...16L}, considerably less than what is required to be associated with PSR\,J1734$-$3333.

\section{Radio spectral index}

Besides the astrometry analysis of the sources using the S-band of the
VLA, we also observed the field in L-band during the 2015 epoch. This
allows for the measurement of the spectral index of the pulsar.

The spectral index for pulsars is defined from a relation between the radio flux density $S$ and the observing frequency $f$, given by $S\propto f^\alpha$.  
Using the 2015 observation in the L band, we measure a flux density for PSR J1734$-$3333 of $710\pm70\,\mu$Jy, centred at $1.5$\,GHz.
Combining this with the S-band ($3.0$\,GHz) 2015 measurement of $344\pm8\,\mu$Jy (Table \ref{tab:src1}) we obtain a spectral index $\alpha=-1.1\pm0.3$.  
The flux density of the pulsar in S-band does not change significantly between 2015 and 2019, so this result is the same if the
S-band measurement from 2019 was used instead. 
A spectral index of $-1.1$ is higher but still consistent with values measured for other normal pulsars, which exhibit a mean between $-1.8$ and $-1.6$ \citepads{1998ApJ...506..863T,2000A&AS..147..195M,2018MNRAS.473.4436J}. 
Thus the spectrum of PSR J1734$-$3333 is somewhat flatter than the spectra of most other radio pulsars. 
Magnetars, on the other hand, when detected in the radio exhibit flat radio spectra \citepads[in general $\alpha<-0.8$ for the magnetars, e.g.][]{2021MNRAS.505.1311H}.

\section{Discusion}
Interferometric observations of PSR J1734$-$3333 were performed on two epochs separated by $4.2$\,yr.  
Astrometric measurements and comparison of both images allow a marginal detection of the proper motion, $\mu_\alpha=10\pm10$\,mas\,yr$^{-1}$ and $\mu_\delta=-29\pm11$\,mas\,yr$^{-1}$, which indicates that the pulsar and SNR G354.8$-$0.8 are not related.  
We also consider a scenario in which this detection is not real and assume an upper limit
on the total proper motion of $\mu < 28$\,mas\,yr$^{-1}$.  
Both scenarios suggest that the pulsar is older than $45$-$100$\,kyr, which can be compared to its characteristic age of only $8.1$\,kyr.

The spectral index that we measure is flatter than what is found for most pulsars and approaches those measured for the six magnetars detected at radio wavelengths. 
Thus not only the rotational properties but also some aspects of the radio emission of PSR J1734$-$3333 resemble magnetar properties.
If PSR J1734$-$3333 displays one day magnetar-like high energy behaviour, then it will be evident that other magnetars also had a past-life as radio pulsars and are much older than what has normally been assumed.
In the likely scenario that magnetars are powered by intense magnetic fields --and assuming that spin-down based measurements can reliably be used to estimate $B$ in this type of objects, what physical mechanism could make high-$B$ pulsars evolve into even-higher-$B$ magnetars?

While the X-ray properties of this pulsar are unremarkable in general, the temperature measured from those observations suggests that it may be hotter than other pulsars of similar $\tau_c$, though not as hot as the magnetars \citepads{2013ApJ...764....1O,2013ApJ...764..180K}.  
If the pulsar was older than $45$\,kyr, then its temperature would be particularly high for its age, thereby supporting the idea that the magnetic field could be evolving quickly in this pulsar \citepads[the amplitude of its surface magnetic dipole component increasing,][]{1996ApJ...458..347M,2011ApJ...741L..13E,2011MNRAS.414.2567H,2015MNRAS.452..845H} and generating additional heat \citepads[e.g.][]{2006MNRAS.371..477K,2012MNRAS.422.2632H,2020ApJ...903...40D}.  
Such evolution could be caused by the emergence of a field that was buried by accretion from a short-lived fallback disc after the supernova explosion \citepads{1989ApJ...346..847C,2013ApJ...770..106B}. 
There are calculations that relate pulsar proper motions with the rate of magnetic field growth, which in turn is related to the braking index \citepads{2013MNRAS.430L..59G}.
For PSR J1734$-$3333 this model could accommodate transverse speeds of up to $\sim500$\,km\,s$^{-1}$, which is somewhat slower than our measurement.
An alternative scenario is that the spin evolution of the pulsar has been driven by accretion from a permanent fallback disc, for which models predict a pulsar age in a similar range to the one we obtain \citepads{2013MNRAS.431.1136C,2014RAA....14...85L}.
In this model magnetars are not powered by the decay of very powerful magnetic fields but by accretion \citepads{2012ApJ...758...98C}. 

More timing observations to monitor the braking index and more sensitive X-ray studies will help to assess the likelihood of the above scenarios and understand the future of PSR J1734$-$3333.

\section{Acknowledgements}
C.M.E acknowledges support by ANID FONDECYT/Regular grants 1171421 and 1211964.
M.V-N. acknowledges support by ANID FONDECYT grant 11191205.
\bibliographystyle{aa}
\bibliography{export-bibtex}



\end{document}